\documentclass[aip,amsmath,amssymb,reprint,]{revtex4-2}
\usepackage{graphicx}
\usepackage{amssymb}
\usepackage{slashed}
\usepackage{dcolumn}
\usepackage{amsmath}
\usepackage{siunitx}
\usepackage{physics}
\usepackage{bm}
\usepackage{color}
\usepackage[colorlinks=true]{hyperref}

\usepackage{mathrsfs}
\makeatletter

\newcommand{\Rmnum}[1]{\expandafter\@slowromancap\romannumeral #1@}
\makeatother

\begin{document}   
	
\title{Near-unity efficiency and photon indistinguishability for the "hourglass" single-photon source using suppression of the background emission}  
	
\author{Benedek Gaál}
\affiliation{DTU Fotonik, Department of Photonics Engineering, Technical University of Denmark, \O rsteds Plads, Building 343, DK-2800 Kongens Lyngby, Denmark}
	
\author{Martin Arentoft Jacobsen}
\affiliation{DTU Fotonik, Department of Photonics Engineering, Technical University of Denmark, \O rsteds Plads, Building 343, DK-2800 Kongens Lyngby, Denmark}
	
\author{Luca Vannucci}
\affiliation{DTU Fotonik, Department of Photonics Engineering, Technical University of Denmark, \O rsteds Plads, Building 343, DK-2800 Kongens Lyngby, Denmark}

\author{Julien Claudon}
\affiliation{Univ. Grenoble Alpes, CEA, Grenoble INP, IRIG, PHELIQS, “Nanophysique et Semiconducteurs” Group, F-38000 Grenoble, France}
\author{Jean-Michel Gérard}
\affiliation{Univ. Grenoble Alpes, CEA, Grenoble INP, IRIG, PHELIQS, “Nanophysique et Semiconducteurs” Group, F-38000 Grenoble, France}
	
\author{Niels Gregersen}
\thanks{Author to whom correspondence should be addressed}
\email{ngre@fotonik.dtu.dk}
\affiliation{DTU Fotonik, Department of Photonics Engineering, Technical University of Denmark, \O rsteds Plads, Building 343, DK-2800 Kongens Lyngby, Denmark}

\date{\today}
	
\begin{abstract}
An on-going challenge within scalable optical quantum information processing is to increase the collection efficiency $\varepsilon$ and the photon indistinguishability $\eta$ of the single-photon source towards unity. Within quantum dot-based sources, the prospect of increasing the product $\varepsilon \eta$ arbitrarily close to unity was recently questioned. In this work, we discuss the influence of the trade-off between efficiency and indistinguishability in the presence of phonon-induced decoherence, and we show that the photonic "hourglass" design allows for improving $\varepsilon \eta$ beyond the predicted maximum for the standard micropillar design subject to this trade-off. This circumvention of the trade-off is possible thanks to control of the spontaneous emission into background radiation modes, and our work highlights the importance of engineering of the background emission in future pursuits of near-unity performance of quantum dot single-photon sources.
\end{abstract}
	
\maketitle 
	
	
Within scalable optical quantum information processing \cite{Pan2012,OBrien2009}, a key component is the single-photon source \cite{Aharonovich2016,Gregersen2013,Gregersen2017} (SPS) producing the single photons used to encode the quantum bits. 
The $N$-photon interference experiment \cite{Wang2019c,Pan2012}, constituting the backbone of optical quantum computing protocols, has a success probability $P$ scaling as $P = (\varepsilon \eta)^N$, where $\varepsilon$ is the source efficiency \cite{Gregersen2013,Gregersen2017} defined as the probability of detecting a photon per trigger and $\eta$ is the indistinguishability of subsequently emitted photons. 
The SPS performance can thus be evaluated in terms of the product $\varepsilon \eta$, and upscaling the multi-photon interference experiment to a large $N$ thus requires increasing $\varepsilon \eta$ arbitrarily close to unity. 
The spontaneous parametric down-conversion process \cite{Kwiat1995} allows straightforward generation of polarization entangled photon pairs and has been the workhorse within quantum light generation for several decades, however the probabilistic nature of the emission process reduces $\varepsilon$ to a few percent. 
 
As alternative, the semiconductor quantum dot \cite{Aharonovich2016,Shields2007} (QD) has emerged as a leading platform for efficient generation of single indistinguishable photons. The spontaneous emission process of the two-level system of the QD allows for deterministic emission of single photons. However, the large refractive index contrast at the semiconductor-air  interface limits $\varepsilon$ to $\sim$ 0.01 for a QD in bulk medium, and it is necessary to place the emitter inside an optical antenna SPS microstructure to guide the light \cite{Gregersen2013,Gregersen2017} towards the collection optics. 
Furthermore, the interaction with the solid-state environment leads to reduced indistinguishability \cite{Kaer2013} of the emitted photons. 
Photon energy fluctuations due to a fluctuating charge environment \cite{Berthelot2006} can be suppressed by introducing electrical contacts \cite{Somaschi2016,Houel2012} and a static electric voltage to stabilize the charge environment.  
A more fundamental decoherence mechanism, however, is the interaction with quantized lattice vibrations, leading to phonon-induced decoherence \cite{Kaer2013,Iles-Smith2017,Gregersen2013,Gregersen2017}. This mechanism produces a phonon sideband comprised of distinguishable photons in the emission spectrum even at 0 K and results in a maximum indistinguishability of $\sim$ 0.9 for an InAs QD in bulk \cite{Kaer2013} 
and of $\sim$ 0.95 for a GaAs QD in bulk \cite{Scholl2019}, due to shorter lifetime. 
While the photons in this sideband can be removed using a narrow spectral filter, this occurs at the cost of efficiency, and the figure of merit $\varepsilon \eta$ is thus not improved by spectral filtering.  

\begin{figure}[t]
	\centering
	\includegraphics[width=8.2cm]{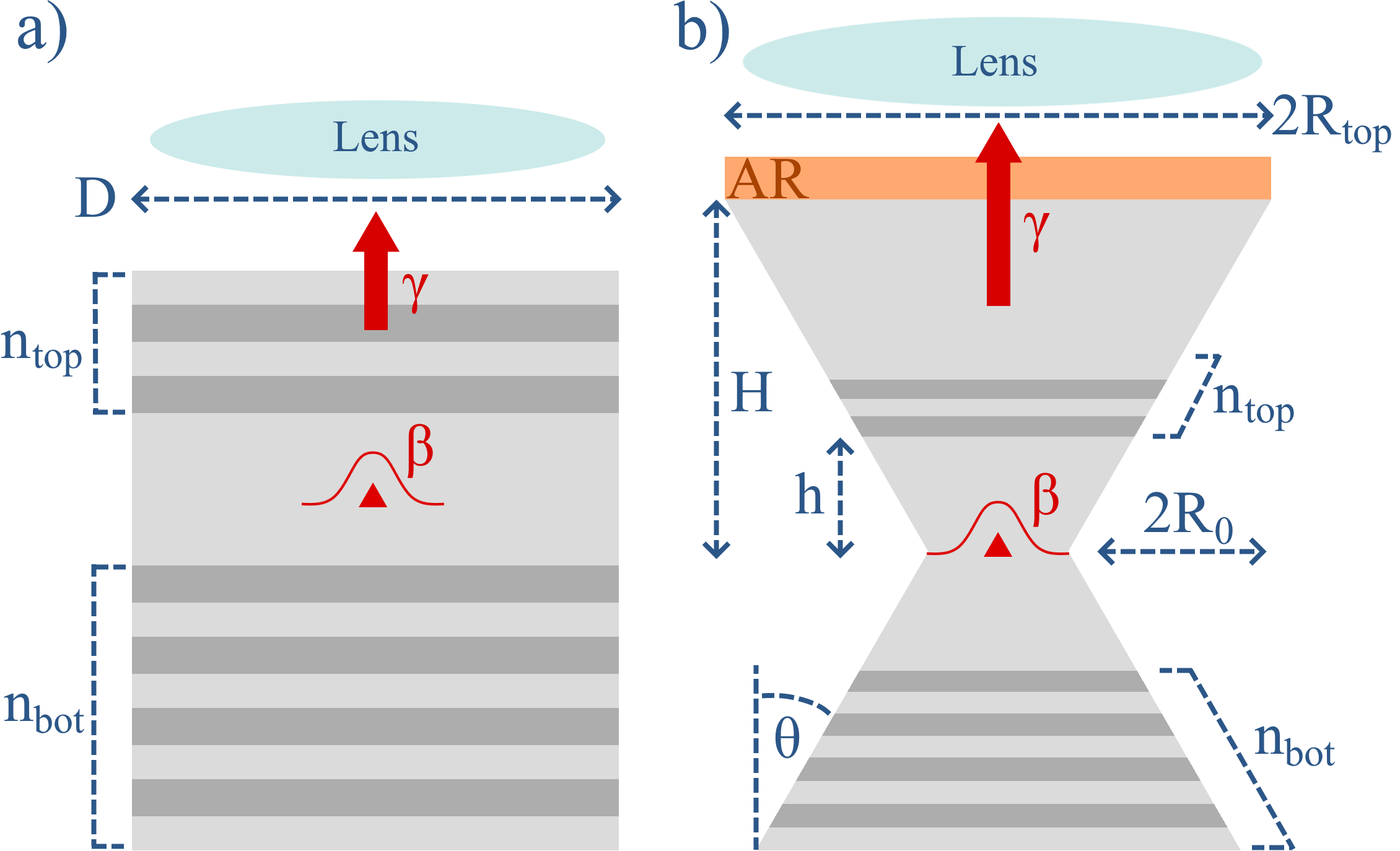}
	\caption{Schematic of (a) the micropillar and (b) the hourglass geometries, where the QD is represented by a red triangle. Both structures feature distributed Bragg reflectors (DBRs) with $n_{\rm top}$ ($n_{\rm bot}$) layer pairs above (below) the QD.}
	\label{fig:geometry}
\end{figure}

Whereas all SPS design approaches require spatial alignment between the QD and the optical mode profile, broadband design strategies, such as the photonic nanowire \cite{Claudon2010,Bleuse2011,Munsch2013, Munsch2013b, Osterkryger2019,Gregersen2016, Gregersen2008, Gregersen2010a,Claudon2013, Nowicki-Bringuier2008, Cadeddu2016, Munsch2012}, the bullseye \cite{Wang2019a,Liu2019}, the microlens \cite{Gschrey2015a}, the photonic crystal waveguide \cite{Arcari2014,MangaRao2007,Lecamp2007b} and the planar dielectric antenna \cite{Lee2011} approach, do not rely on resonant effects. For example, the photonic nanowire design instead exploits a dielectric screening effect \cite{Bleuse2011} to suppress emission into radiation modes and ensure preferential coupling to the fundamental waveguide mode, 
whereas the photonic crystal waveguide \cite{Arcari2014} combines dielectric screening with the slow light effect \cite{MangaRao2007,Lecamp2007b} near the photonic band edge to also enhance emission into the waveguide mode.
Thus, the strength of a broadband strategy is that careful spectral alignment between a cavity resonance and the QD emission line is unnecessary. However, a drawback is the lack of a mechanism to suppress the phonon sideband \cite{Iles-Smith2017}. 
In contrast, narrowband SPS design strategies including the micropillar \cite{Somaschi2016, Ding2016, Wang2019a, Wang2020_PRB_Biying, Wang2021, Huber2020, Ates2009, Gayral1998, Gur2021} shown in Fig.\ \ref{fig:geometry}(a) and the open cavity geometry \cite{Tomm2021} rely on the resonant cavity quantum electrodynamics (CQED) effect to funnel emission \cite{Gregersen2013,Gregersen2017} into a well-defined cavity mode at the resonance wavelength $\lambda _{\rm C}$. The CQED design approach requires a spectral alignment between the QD and the cavity mode, and with typical $Q$ factors \cite{Somaschi2016,Ding2016,Wang2019a} above 5000, a tuning mechanism for the QD line is generally needed. However, an advantage of the narrowband approach is the efficient funneling of photons into the cavity mode leading to a suppression of the phonon sideband \cite{Iles-Smith2017,Wang2020_PRB_Biying} and measured indistinguishability \cite{Ding2016,Somaschi2016} of $\sim$ 0.99 without spectral filtering.

In both design approaches, the physics of the coupling efficiency $\varepsilon$ is often analyzed using a single-mode model \cite{Gerard1998,Wang2020_PRB_Biying,Osterkryger2019,Gregersen2013}, where the single-mode efficiency $\varepsilon _{\rm s}$ is given as the product $\varepsilon _{\rm s} = \beta \gamma$ (cf.\ Fig.\ \ref{fig:geometry}). Here, the spontaneous emission $\beta$ factor is the fraction of light coupled to the fundamental cavity mode $\beta = \Gamma_{\rm C} / \Gamma_{\rm T}$, with $\Gamma_{\rm C}$ and $\Gamma_{\rm T}$ being the emission rate into the cavity mode and the total rate, respectively, and the transmission $\gamma$ is the fraction of power in the cavity detected by the collection optics for a specific numerical aperture (NA) taking into account an overlap with a Gaussian profile \cite{Munsch2013,Munsch2013b}.
The emission rate $\Gamma_{\rm C}$ into the cavity normalized to that of a bulk medium $\Gamma_{\rm Bulk}$ is given by the Purcell factor $F_{\rm p} = \Gamma_{\rm C}/\Gamma_{\rm Bulk} = \frac{3}{4 \pi^2} \frac{Q}{V_{\rm n}}$, where $Q$ is the cavity quality factor,  $V_{\rm n}$ is the mode volume in units of $(\lambda/n)^3$ and $n$ is the cavity refractive index. 
In terms of the Purcell factor, $\beta$ can be written as
 \begin{align}
	\beta = \frac{\Gamma_{\rm C}}{\Gamma_{\rm C} +  \Gamma_{\rm B}} = \frac{F_{\rm p} }{F_{\rm p}  + \Gamma_{\rm B} / \Gamma_{\rm Bulk}},
	\label{beta_eq}
\end{align}
where $\Gamma_{\rm B} = \Gamma_{\rm T} - \Gamma_{\rm C}$ is the total background emission rate. From Eq.\ \eqref{beta_eq}, it would appear that $\beta$ and in turn the efficiency can be optimized simply by increasing $F_{\rm p}$. Furthermore, increasing the Purcell factor by enhancing the $Q$ factor brings the additional benefit of improving the indistinguishability in the presence of phonon-induced decoherence by funneling the spontaneous emission through a spectrally narrow cavity mode \cite{Iles-Smith2017, Wang2020_PRB_Biying}, as discussed above, without reducing the efficiency.
Thus, the acknowledged optimization strategy of the CQED-based SPS design both in terms of efficiency and indistinguishability has been to increase the Purcell factor. 

However, it was recently shown that this simple optimization paradigm only works \cite{Iles-Smith2017} in the weak coupling regime: As the Purcell factor increases, the strong coupling regime is reached leading to the generation of two hybrid polariton states, whose energy splitting (of the order of $\sim$ 0.1 meV) is smaller than the acoustic phonon cutoff energy (1--2 meV). Therefore, phonon modes are available to mediate transitions between the polariton states, resulting in decoherence in the emission process \cite{Denning2020_strongcoupling}. Thus increasing the Purcell factor beyond a certain point will actually reduce the indistinguishability, and the standard CQED design approach thus features an inherent trade-off \cite{Iles-Smith2017} between achievable efficiency and indistinguishability. Subject to this trade-off, the micropillar SPS design was recently numerically optimized \cite{Wang2020_PRB_Biying} in terms of efficiency and indistinguishability with a maximum product $\varepsilon \eta$ obtained for $\varepsilon$ = 0.95 and $\eta$ = 0.997. While these figures of merit have yet to be demonstrated experimentally, the question remains: Is the trade-off of fundamental nature, or can the performance be improved further using unconventional design strategies?

One way of avoiding phonon-induced decoherence is obviously to suppress the interaction with the phonons. Direct phononic engineering, where phonons are suppressed using a phononic bandgap, generally requires nanofabrication on a length scale smaller than what is possible today. However, another option is to suppress the photon emission taking place via phonon-assisted emission channels. The idea here is to stay within the weak coupling regime and optimize the spontaneous emission $\beta$ factor in \eqref{beta_eq}, not by increasing the Purcell factor, but instead by reducing the emission $\Gamma_{\rm B}$ into background modes. Indeed, the broadband photonic nanowire \cite{Claudon2010,Bleuse2011,Munsch2013, Munsch2013b, Osterkryger2019,Gregersen2016, Gregersen2008, Gregersen2010a,Claudon2013, Nowicki-Bringuier2008, Cadeddu2016, Munsch2012} design approach relies on this exact concept, where dielectric screening \cite{Bleuse2011} is used to suppress the background (radiation) modes. 

In the remainder of this Perspectives Letter, we address the above question by considering a specific SPS design based on the recently proposed hybrid photonic nanowire-micropillar "hourglass" design \cite{Osterkryger2019} shown in Fig.\ \ref{fig:geometry}(b). The hourglass features a narrow waist at the position of the QD, enabling suppression of the background emission using dielectric screening, as well as tapers featuring distributed Bragg reflectors (DBRs), allowing for Purcell enhancement of the spontaneous emission. 
Using the example of the hourglass, we show how careful engineering of the background emission may indeed pave the way for future increase of the SPS figures of merit towards unity. As a first step down this avenue, we demonstrate that the figures of merit can be improved beyond the limitation imposed by the trade-off for the standard micropillar SPS design.

	
We perform optical simulations using a Fourier Modal Method \cite{gur2020open,NUMERICALMETHOD2014,DTU_DCC_resource} (FMM) with a true open boundary condition to determine the efficiency. The QD is modeled as a classical point dipole with in-plane dipole orientation and harmonic oscillation frequency $\omega$. 
The previously defined spontaneous emission rates $\Gamma_{\rm X}$ (X = C, B, T) are then determined as $\Gamma_{\rm X}/\Gamma_{\rm Bulk} = P_{\rm X}/P_{\rm Bulk}$ \cite{novotny2012principles}, where $P_{\rm X}$ are the corresponding classical powers emitted by the dipole and $\Gamma_{\rm Bulk}$ ($P_{\rm Bulk}$) is the spontaneous emission rate (power) in a bulk medium. 
For our single-mode model with $\varepsilon _{\rm s} = \beta \gamma$, the transmission $\gamma$ from the cavity mode is computed as $\gamma = P_{\rm Lens,C} / P_{\rm C}$, where $P_{\rm Lens,C}$ is the power coupled to the lens from the cavity mode alone and $P_{\rm C}$ is the power emitted into the cavity mode. The calculation of $P_{\rm Lens,C}$ includes an overlap integral with a Gaussian profile \cite{Munsch2013,Munsch2013b} to model coupling to the fundamental mode of a single-mode fiber (SMF). 
Similarly, the efficiency $\varepsilon$ computed using the full model is determined as $\varepsilon = P_{\rm Lens} / P_{\rm T}$, where $P_{\rm Lens}$ is the output power coupled to the lens, again taking into account an overlap with a Gaussian mode.

The indistinguishability is calculated using a Born-Markov master equation in the polaron frame including the effects of the phonon environment \cite{Wang2020_PRB_Biying,Stephen2015,Iles-Smith2017,Denning2018Cavity-waveguideSources,wilson2002quantum} discussed in detail in Ref.~\onlinecite{Wang2020_PRB_Biying}. 
To ensure a fair comparison with the micropillar optimization \cite{Wang2020_PRB_Biying}, we use the same parameters for the phonon environment as in Ref.~\onlinecite{Wang2020_PRB_Biying}. 
In the high-$\beta$ regime, where light from the QD reaching the collection optics is emitted almost exclusively via the cavity mode, 
the indistinguishability depends only on three parameters from the optical simulations, namely the cavity escape rate $\kappa = \omega / Q$, the cavity-QD coupling strength $g$ proportional to $1 / \sqrt{V_{\rm n}}$ and the background emission rate $\Gamma_{\rm B}$. 
In this Letter, instead of performing a full parameter sweep, we propose an hourglass design with a similar value of $g$ as that for the optimum micropillar design, \cite{Wang2020_PRB_Biying} and we investigate the influence of the dielectric screening effect on the performance.


Similar to the micropillar, the hourglass consists of an InAs QD in a vertical GaAs cavity surrounded by distributed Bragg reflectors (DBRs) with $n_{\rm top}$ ($n_{\rm bot}$) layer pairs in the top (bottom) DBR.
However, the center cavity radius is chosen as $R_0$~=~114~nm to suppress the background emission rate using the dielectric screening effect \cite{Gregersen2013,Bleuse2011}, such that $\Gamma_{\rm B} / \Gamma_{\rm Bulk} \sim$~0.05 at our design wavelength $\lambda$~=~925~nm. 
The hourglass then features inverted "trumpet" tapers, where the top taper is used to reduce the output beam divergence and ensure a good coupling to the collection optics and the bottom taper is needed to obtain high bottom DBR reflectivity as discussed in detail in Ref.\ \onlinecite{Osterkryger2019}. 
In this Letter, we consider a symmetric structure with identical sidewall angle $\theta$ and identical QD-DBR separation height $h$ in both top and bottom taper sections.
The DBRs consist of alternating layers of Al$_{0.85}$Ga$_{0.15}$As and GaAs with refractive indices\cite{Gehrsitz2000} $n_{\rm AlGaAs}=2.9895$ and $n_{\rm GaAs}=3.4788$ for $\lambda$~=~925~nm and $T$~=~4~K, and we use diameter-dependent thicknesses of the DBR layers using the procedure outlined in Appendix C of Ref.~\onlinecite{Gregersen2016}.
Above the top DBR, the hourglass features an additional homogeneous top taper section followed by an anti-reflection (AR) coating with refractive index $n_{\rm AR}=\sqrt{n_{\rm GaAs}}$ and thickness given by $t_{\rm AR}=\lambda/4n_{\rm eff, AR}$ to prevent the formation of a second cavity in this top taper section. 

\begin{figure}[tb]
	\centering
	\includegraphics[width=8.2cm]{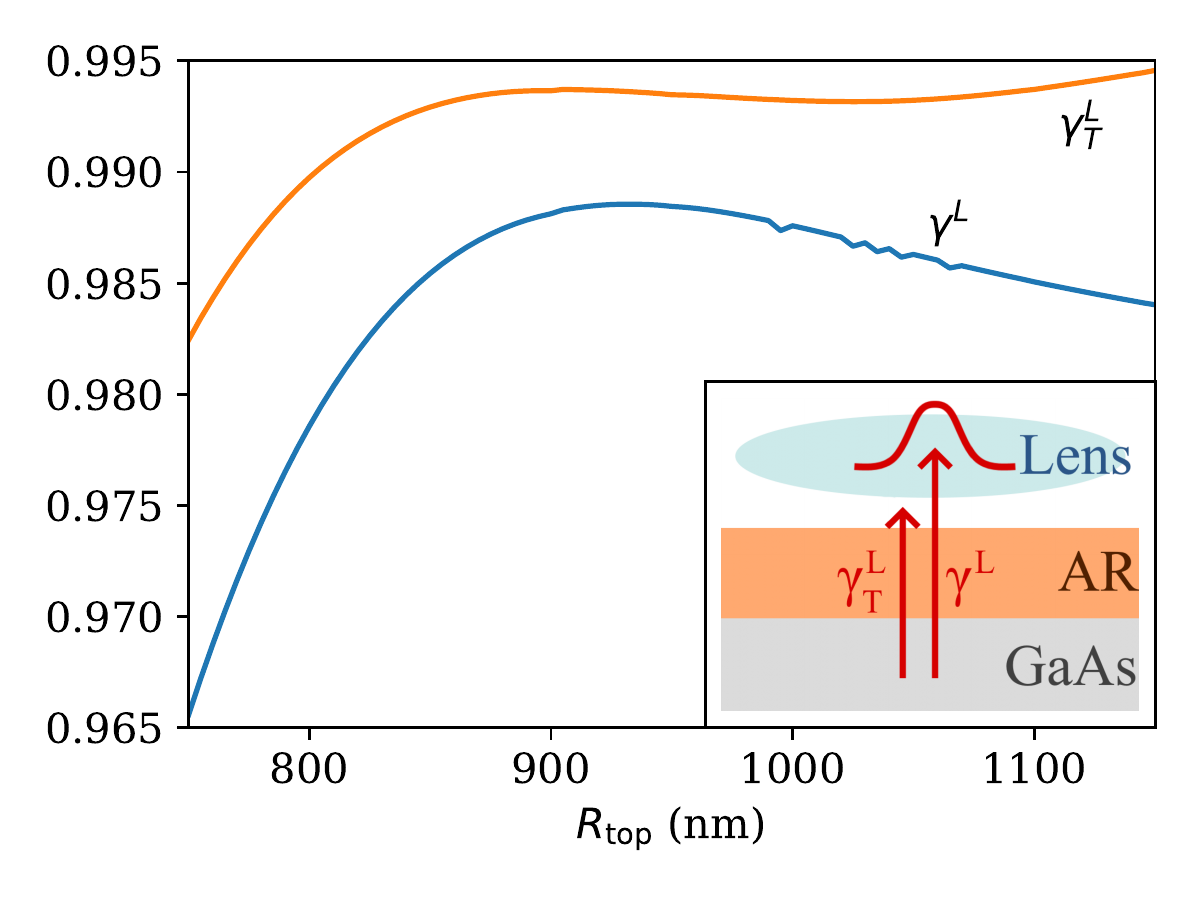}
	\caption{Transmission $\gamma ^{\rm L}$ including the Gaussian overlap and total transmission $\gamma ^{\rm L}_{\rm T}$ for the fundamental HE$_{11}$ mode as function of top radius $R_{\rm top}$ for a 0.82 NA collection lens. The transmission is computed for the three-layer structure shown in the inset.}
	\label{fig:gaussian}
\end{figure}

To determine a top taper geometry allowing for high collection efficiency, we write the transmission $\gamma$ as the product $\gamma = \gamma^{\rm L} {\rm T_{11}}$, where T$_{11}$ is the power transmission of the fundamental HE$_{11}$ mode from the QD to the top of the GaAs taper in the absence of a DBR and $\gamma^{\rm L}$ is the HE$_{11}$ transmission from the top of the GaAs taper to the collection lens (cf.\ Fig.~\ref{fig:gaussian}, inset) taking into account the Gaussian overlap.

We present the transmission $\gamma^{\rm L}$ as well as the total transmission $\gamma^{\rm L}_{\rm T}$ without the Gaussian overlap in Fig.~\ref{fig:gaussian} as function of $R_{\rm top}$. For decreasing values of $R_{\rm top}$ below $\sim$~900 nm, the output beam divergence increases \cite{Gregersen2008} and both transmissions are reduced. For increasing $R_{\rm top}$, the total transmission $\gamma^{\rm L}_{\rm T}$ increases towards 1 as the output beam divergence decreases. Importantly, we notice that the transmission $\gamma^{\rm L}$ taking into account the Gaussian overlap does not increase towards unity with $R_{\rm top}$ but instead takes a maximum value of $\gamma^{\rm L}_g=0.988$ at $R_{\rm top}=\SI{930}{nm}$. This ceiling occurs due to imperfect mode overlap of the strongly confined mode in the GaAs waveguide (Bessel function profile) and the Gaussian profile resulting from weak confinement in a SMF. In the following, we choose $R_{\rm top}=\SI{930}{nm}$ corresponding to maximum transmission to the lens. 

\begin{figure}[tb]
	\centering
	\includegraphics[width=7.6cm]{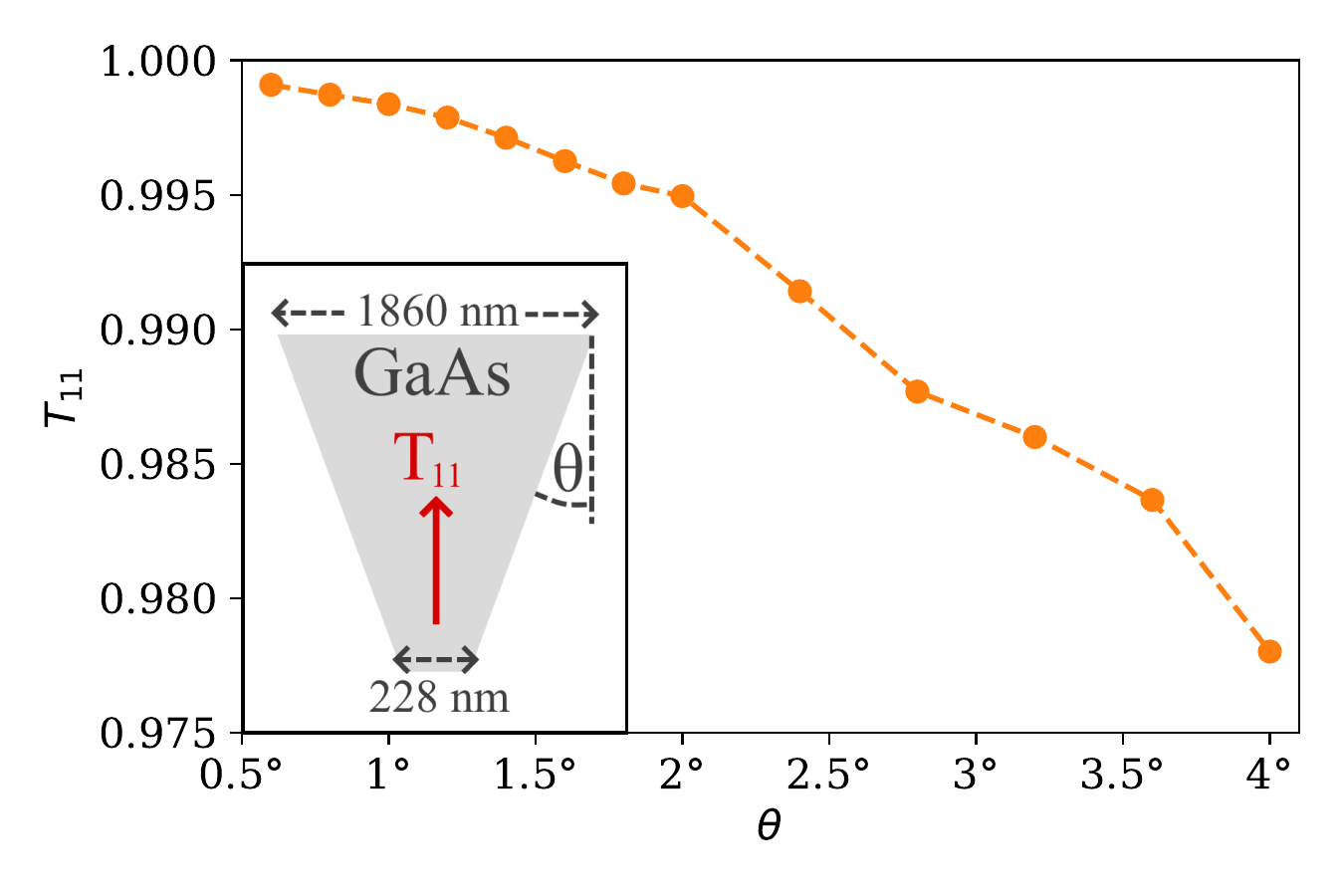}
	\caption{The power transmission T$_{11}$ of the $\rm HE_{11}$ mode through the bare top taper shown in the inset as function of the sidewall angle $\theta$.}
	\label{fig:taper}
\end{figure}

Having fixed $R_0$ and $R_{\rm top}$, we proceed to determine a taper geometry allowing for an adiabatic expansion \cite{Gregersen2010a} of the fundamental mode suppressing coupling to higher orders modes and allowing for a transmission $T_{11} \rightarrow$~1. In this work, we consider a linear taper with sidewall angle $\theta$. The computed transmission as function of $\theta$ is presented in Fig.~\ref{fig:taper} illustrating how a high transmission is obtained at the expense of a small sidewall angle, in turn leading to a tall structure. Weak oscillations for large $\theta$ due to higher-order mode coupling \cite{Gregersen2010a} are also observed. We choose a sidewall angle of $\theta=0.8^{\circ}$ leading to a total top taper height H of 58.5 $\mu$m and a transmission $T_{11}$ = 0.9987. 

We now consider the full hourglass structure including the bottom taper with the DBR, where we set the number of bottom DBR layer pairs $n_{\rm bot}$~=~46 to avoid leakage of light into the substrate. We then fix the QD-DBR separation height to $h$~=~24142~nm such that the QD is placed at a field antinode and such that the normalized mode volume $V_{\rm n}$ will take a value of $\approx$ 28, a value which leads to the optimum $\epsilon \eta$ figure of merit for the micropillar SPS design \cite{Wang2020_PRB_Biying}. 

\begin{figure}
	\centering
	\includegraphics[width=8.2cm]{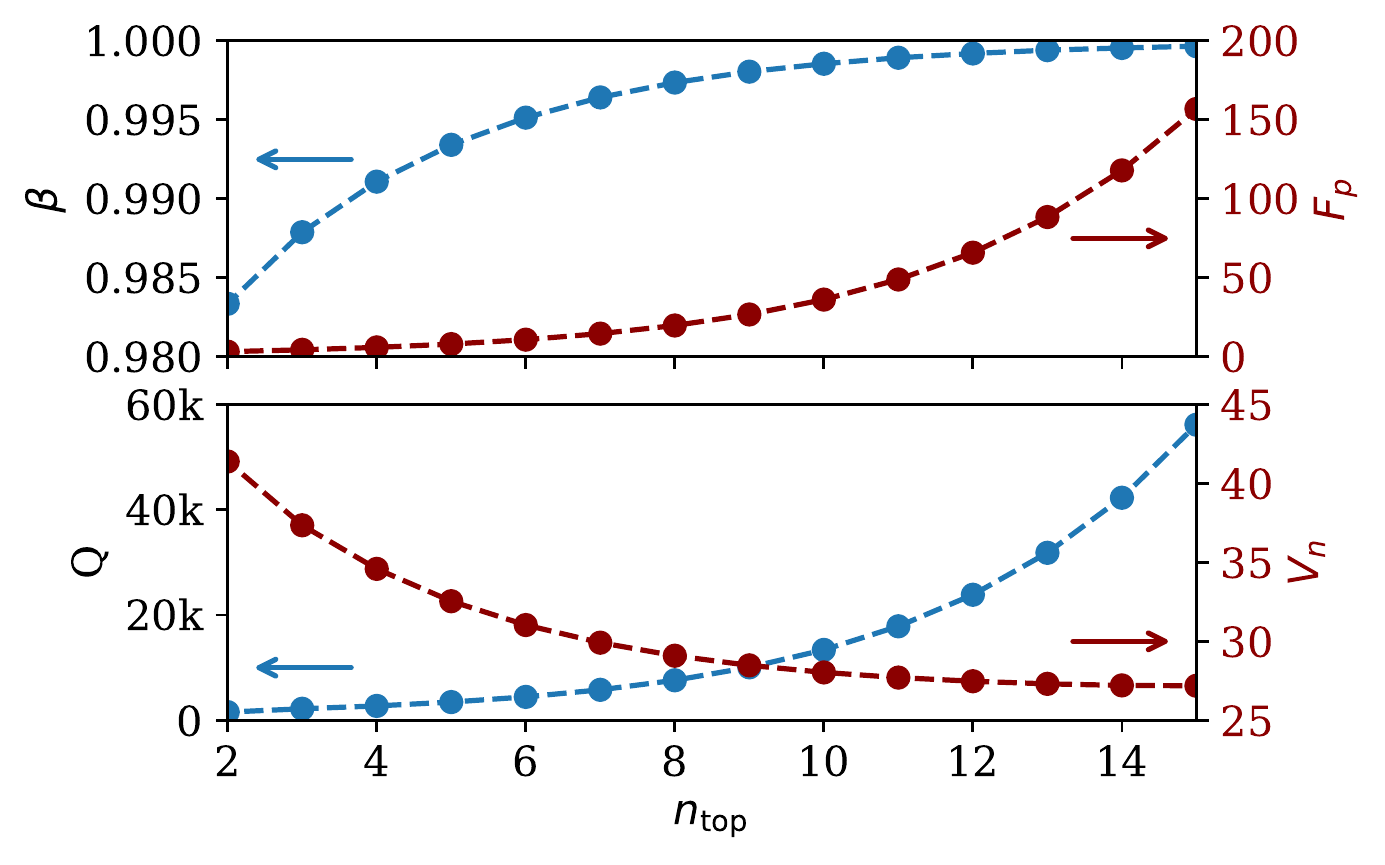}
	\caption{The $\beta$ factor, the Purcell factor $F_{\rm p}$, the $Q$ factor and the mode volume $V_{\rm n}$ of the hourglass geometry as a function of top DBR layer pairs $n_{\rm top}$ with $n_{\rm bot}$~=~46.}
	\label{fig:QV}
\end{figure}

The computed spontaneous emission $\beta$ factor, Purcell factor $F_{\rm p}$, $Q$ factor and mode volume $V_{\rm n}$ are presented in Fig.~\ref{fig:QV} as function of the number of top layer pairs $n_{\rm top}$.
As expected, the $Q$ factor increases rapidly with $n_{\rm top}$. For $n_{\rm top} \geq$~8, the normalized mode volume $V_{\rm n}$ decreases slowly below 30 leading to an increase in $F_{\rm p}$ with $n_{\rm top}$ roughly proportional to that in $Q$ with $F_{\rm p} \approx$ 150 for $n_{\rm top}$~=~15.
However, we observe that a $\beta$ factor above 0.98 is obtained for all values of $n_{\rm top}$, even for modest values of the Purcell factor. This excellent coupling is a result of the nanowire effect of suppressing the background emission \cite{Bleuse2011} in the low-diameter regime and represents a major asset of the hourglass SPS design. As for the $Q$ factor, we note that the optimal photon indistinguishability in Ref.~\onlinecite{Wang2020_PRB_Biying} was obtained for a $Q\sim$~30,000, and we observe that this value is reached for $n_{\rm top}$~=~11. 

\begin{figure}[tb]
	\centering
	\includegraphics[width=7cm]{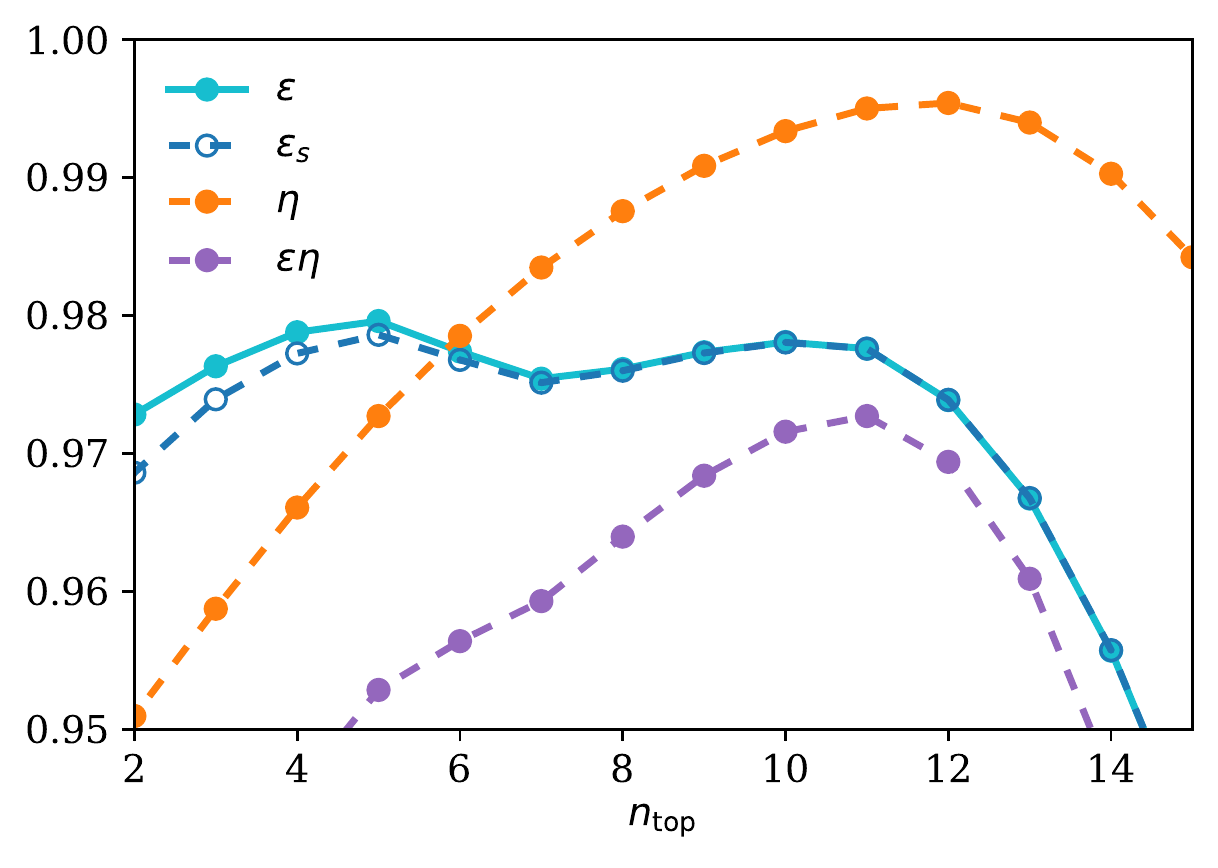}
	\caption{The efficiency $\varepsilon$ of the full model, the single-mode efficiency $\varepsilon _{\rm s}$, the indistinguishability $\eta$ and the product $\varepsilon \eta$ of the hourglass design as function of $n_{\rm top}$.}
	\label{fig:eps_eta}
\end{figure}

We then present the calculated hourglass efficiency and the photon indistinguishability in Fig.~\ref{fig:eps_eta} as function of $n_{\rm top}$. The predictions for the efficiency from the full model $\varepsilon$ and the single-mode model $\varepsilon_{\rm s}$ agree well, confirming that the outcoupling of light to the collection lens is mediated almost exclusively by the cavity mode. 
The small discrepancy for the smaller cavities occurs due to interaction with radiation modes as discussed in Ref.\ \onlinecite{Osterkryger2019}.
We notice that, unlike the micropillar design, the maximum efficiency of 0.98 is obtained for a modest top mirror reflectivity highlighting the fact that the hourglass design does not rely on significant Purcell enhancement to ensure high efficiency. 
A small oscillation is observed between $n_{\rm top}$~=~5 and 11 occurring due to a weak cavity coupling effect with the taper section above the DBR, whereas for $n_{\rm top} \geq$~5 the efficiency drops uniformly due to increasing leakage of light into the substrate. 
The indistinguishability increases with top mirror reflectivity with a maximum value of  $\eta$~=~0.995 at $n_{\rm top}$~=~12, after which the penetration into the strong-coupling regime \cite{Iles-Smith2017} leads to reduction in $\eta$.
We then consider the product of efficiency and indistinguishability and observe that it attains a maximum value of $\varepsilon\eta=0.973$ at $n_{\rm top}=11$ representing a clear improvement compared to maximum value for the micropillar geometry of $\varepsilon\eta$ = 0.95 from Ref. \onlinecite{Wang2020_PRB_Biying}.
These results for the hourglass design confirm that the trade-off \cite{Iles-Smith2017} between efficiency and indistinguishability for the cavity-based SPS is not fundamental but can indeed be overcome by careful engineering of the emission into background radiation modes.

Our analysis demonstrates that the combination of dielectric screening and Purcell enhancement is an excellent strategy to attain a near-unity beta factor.
The cavity element of the hourglass design operates at an almost ideal level, where $\sim$~99.9~\% of the emitted light is coupled to the fundamental cavity mode.
On the other hand, the photon indistinguishability is not improved compared to the micropillar \cite{Wang2020_PRB_Biying}. Its value is influenced by $g$, $\kappa$ and $\Gamma_{\rm B}$, and by construction, our values of $g$ and $\kappa$ are in good agreement with those obtained \cite{Wang2020_PRB_Biying} for the micropillar design. While $\Gamma_{\rm B}$ is reduced by an order of magnitude \cite{Wang2020_PRB_Biying,Wang2021}, the spontaneous emission rate is dominated by the Purcell enhanced cavity emission, and the reduction in $\Gamma_{\rm B}$ does not lead to a substantial improvement in the indistinguishability. 

In this Perspectives letter, the improvement in the overall $\varepsilon \eta$ figure of merit is thus brought about by the ability to efficiently funnel light into the cavity mode while staying in the weak coupling regime and not by a reduction in phonon-assisted background emission.
However, for an SPS design such as the hourglass, where the background emission rate is controlled,  suppression of phonon-assisted emission \textit{via the cavity} and a corresponding improvement in $\eta$ can be introduced by increasing $Q$ \cite{Iles-Smith2017} while simultaneously increasing the mode volume to avoid the strong coupling regime.
While we leave such optimization of the hourglass SPS design to a follow-up work, we remark that control of the background emission within SPS engineering represents a new avenue to be explored for highly efficient emission of single indistinguishable photons. Whereas the photonic nanowire effect \cite{Claudon2010,Gregersen2013,Gregersen2017,Bleuse2011,Claudon2013,Nowicki-Bringuier2008} and proper choice of pillar diameter \cite{Wang2021} represent initial steps along this avenue, further ideas such as photonic bandgap engineering using e.g.\ circular Bragg grating rings surrounding the main waveguide remain to be explored. 

While the concept of engineering of the background emission is of interest for all QD-based SPS designs, 
we now consider prospects for increasing the efficiency and indistinguishability of specifically the hourglass design further towards unity. A near-unity beta factor means that the bottleneck preventing unity coupling efficiency is now moved from the cavity to other elements. In the following, we discuss limitations 
associated with (i) the coupling to a SMF, (ii) charge noise and (iii) the performance of the QD excitation scheme.

Whereas the total transmission from the HE$_{11}$ mode to a collection lens can be increased towards unity simply by increasing the top radius $R_{\rm top}$, cf.\ Fig.\ \ref{fig:gaussian}, the coupling to a Gaussian profile\cite{Gaussian_overlap} of a SMF is limited to 0.988. This limitation could be overcome using butt coupling \cite{Cadeddu2016,Howe1996} of an SMF to the top surface of the hourglass. By tapering of the SMF, the radius of the glass waveguide at the contact point can be chosen to match $R_{\rm top}$ such that the mode overlap considered is between strongly confined (Bessel function profiles) modes both in the GaAs and the glass waveguide. 
An associated practical challenge is the significant height H of 58.5 $\mu m$ of the top taper required for a $T_{11}$ of 0.9987. Since the large value of 930 nm of $R_{\rm top}$ for coupling to a 0.82 NA lens is no longer needed, the butt coupling scheme could simultaneously allow for a reduction of the height of the top taper. We note that non-linear taper shapes, e.g.\ a horn shape \cite{Burns1977}, could also be considered to decrease the taper height.

A disadvantage of the center cavity radius $R_0$ of 114~nm is the close proximity of the QD to surface defect charge states resulting in charge noise and reduced indistinguishability. Here, surface passivation \cite{Liu2018} and electrical contacts \cite{Somaschi2016,Houel2012} may be implemented to stabilize the charge environment.

A third frequently overlooked source of loss arises from the QD excitation strategy. In the present work, it is assumed that the emitter is initialized in the excited state with 100~\% probability using a $\pi$ pulse. However, for vertical resonant excitation, the cross-polarization filtering needed to isolate the emitted photons from the excitation laser \cite{Somaschi2016, Ding2016} reduces the outcoupling efficiency by a factor of 2. While side excitation \cite{Ates2009, Huber2020} is possible, the combination of high collection efficiency and clean emission remains a challenge. 

Now, for the vertical excitation, the introduction of an elliptical cross-section enables polarization control by exploiting spectral separation of the two otherwise degenerate polarization modes \cite{Gayral1998}. For the micropillar SPS, this has allowed for polarized collection efficiency of 0.60 demonstrated \cite{Wang2019a} and 0.90 predicted \cite{Gur2021} to a first lens, and more recently 0.57 demonstrated \cite{Tomm2021} coupling from an open cavity SPS to a SMF. In the low-diameter photonic nanowire regime, selective deconfinement of the undesired polarization using an elliptical cross-section has been demonstrated \cite{Munsch2012} with a polarization control of 0.95 for an unstructured nanowire. For the hourglass geometry, selective deconfinement using an elliptical cross-section at the position of the QD combined with Purcell enhancement should allow for even better polarization control and will be explored in a follow-up work. Even so, a price to pay in all elliptical cross-section schemes is a significant increase in the pump power needed to achieve 100~\% excitation of the excited state. 

Electrical driving has been considered as a practical solution to remove the need for an excitation laser. In this approach, the QD is populated with carriers injected from electrical contacts, with a performance that is on-par with the optical pumping \cite{Bockler2008, Gregersen2010a} and measured extraction efficiency of 0.61 \cite{Schlehahn2016}.
However, the resulting indistinguishability has remained significantly low (in the range 0.41--0.64 \cite{Patel2010, Schlehahn2016}) due to an unavoidable time-jitter effect \cite{Kiraz2004, Kaer2013}.
Non-resonant excitation schemes represent alternative strategies for eliminating the cross-polarization filtering and overcoming the factor of 2 barrier. Schemes based on a non-resonant phonon-assisted excitation \cite{Glassl2013, Ardelt2014, Quilter2015, Thomas2021} have been suggested with predicted \cite{Gustin2020} maximum pumping efficiency up to 0.9 and demonstrated \cite{Thomas2021} indistinguishability of 0.91.
Two-photon excitation of the bi-exciton level \cite{Schweickert2018}, rapidly followed by stimulated emission from the bi-exciton to the exciton level \cite{Sbresny2022, Wei2022, Yan2022}, has recently shown to provide slightly higher measured indistinguishability (0.93), together with in-fiber efficiency up to 0.51 \cite{Wei2022}.
The dichromatic pumping scheme \cite{He2019, Koong2021, Bracht2021, Karli2022} has allowed for even higher demonstrated \cite{He2019} indistinguishability of 0.97 at a pumping efficiency of 0.87, and very recently the pumping efficiency reached 0.97 in a demonstration \cite{Karli2022} based on a red-detuned dichromatic scheme.

While the figures of merit for the QD excitation strategies above represent significant improvements to resonant excitation with a cross-polarization setup, they still introduce loss mechanisms comparable to or dominating over the predicted $\varepsilon \eta$ figure of merit of 0.973 (0.95) for the hourglass (micropillar) SPS. The excitation scheme thus represents a potential bottleneck in the pursuit of ideal $\varepsilon \eta$~=~1 SPS performance, and future SPS engineering should carefully take into account requirements of the excitation scheme.


In conclusion, we have considered challenges in increasing the efficiency and the photon indistinguishability of the QD-based SPS towards unity. A fundamental challenge is the trade-off between efficiency $\varepsilon$ and indistinguishability $\eta$ within the CQED design scheme in the presence of phonon-induced decoherence. We have demonstrated that this trade-off can indeed be circumvented by suppression of the spontaneous emission into background radiation modes. Whereas the product $\varepsilon \eta$ for the micropillar design is limited to 0.95 \cite{Wang2020_PRB_Biying}, we show that hourglass design allows increasing $\varepsilon \eta$ to 0.973. This occurs due to an ultra-high spontaneous emission $\beta$ factor above 0.997 enabled by a suppression of the background emission due to the photonic nanowire effect. In the pursuit of QD-based SPSs featuring near-unity efficiency and photon indistinguishability, engineering of the background spontaneous emission rate thus represents a new avenue to be explored.
\\ \\
This work is funded by the European Research Council (ERC-CoG "UNITY", grant 865230) and by the Independent Research Fund Denmark (Grant No. DFF-9041-00046B). We acknowledge fruitful discussions with Emil Vosmar Denning.
	
\section*{Data availability statement}	

The data that support the findings of this study are available from the corresponding author upon reasonable request.
	
\bibliography{refs}

\end{document}